\documentclass[preprint]{IEEEtran}

\usepackage{amsmath,amsfonts}
\usepackage{graphicx}
\usepackage{cite}
\usepackage{url}
\usepackage[
colorlinks=true,
linkcolor=blue,
citecolor=blue,
urlcolor=blue
]{hyperref}

\usepackage[caption=false,font=normalsize,labelfont=sf,textfont=sf]{subfig}
\usepackage{booktabs}
\usepackage{multirow}
\usepackage{xcolor}
\usepackage{colortbl}
\usepackage{threeparttable}
\usepackage{adjustbox}
\usepackage{longtable}
\usepackage{algorithm}
\usepackage{algpseudocode}
\usepackage[utf8]{inputenc} 
\usepackage{textgreek}
\DeclareUnicodeCharacter{03B2}{\ensuremath{\beta}}

\usepackage{enumitem}
\usepackage{pdflscape}
\usepackage{tcolorbox}
\tcbuselibrary{skins, breakable}

\definecolor{sigrow}{RGB}{220,240,232}
\definecolor{warnrow}{RGB}{255,245,215}
\definecolor{gapwarn}{RGB}{255,200,200}

\newcommand{\tierM}{\cellcolor{yellow!25}\textbf{M}}
\newcommand{\tierW}{\cellcolor{orange!25}\textbf{W}}

\newtcolorbox{insightbox}[1][]{
	enhanced,
	breakable,
	colback=teal!6!white,
	colframe=teal!50!black,
	colbacktitle=teal!50!black,
	coltitle=white,
	fonttitle=\bfseries\small,
	title={Key insight},
	arc=4pt,
	boxrule=0.6pt,
	left=6pt, right=6pt, top=4pt, bottom=4pt,
	#1
}
\usepackage{xcolor}
\definecolor{cliffLarge}{RGB}{198,239,206}   
\definecolor{cliffMedium}{RGB}{255,243,176}  
\definecolor{cliffSmall}{RGB}{255,255,255}

\renewcommand{\thesection}{\Roman{section}}
\renewcommand{\thetable}{\Roman{table}}
\renewcommand{\thefigure}{\Roman{figure}}

\begin{document}

\title{Multi-Modal Machine Learning for Population- and Subject-Specific
lncRNA-Type 2 Diabetes Association Analysis}

\author{
	Ashwani Siwach,
	Sanjeev Narayan Sharma,
	and Sunil Datt Sharma
	
	\IEEEcompsocitemizethanks{
		\IEEEcompsocthanksitem Ashwani Siwach and Sanjeev Narayan Sharma are with the Department of Electronics and Communication Engineering, IIITDM Jabalpur, India.\protect\\
		E-mail: ashwanisiwach132003@gmail.com
		\IEEEcompsocthanksitem Sunil Datt Sharma is with the Department of Electronics and Communication Engineering, Central University of Jammu, India.
	}
	\thanks{Corresponding author: Ashwani Siwach.}
}

\maketitle

\begin{abstract}
	Long non-coding RNAs (lncRNAs) are emerging regulatory molecules implicated in chronic disease pathogenesis, including Type 2 Diabetes Mellitus (T2D). We investigated ten literature reported lncRNAs associated with T2D: MALAT1, MEG3, MIAT, ANRIL, GAS5, KCNQ1OT1, H19, BCYRN1, XIST, and HOTAIR across two independent population-based RNA-seq cohorts. Single-omics approaches provide an incomplete view of disease biology, therefore, an integrative multi-feature framework was developed, extracting expression, secondary-structure, and sequence features for each lncRNA. Eight machine learning (ML) classifiers were evaluated under stratified k-fold, leave-one-out cross-validation (LOOCV), and repeated hold-out schemes to ensure robust performance estimation. SHAP analysis was applied for subject-level association interpretation. In one cohort, GAS5 and XIST expression features, along with GAS5, MEG3, and ANRIL sequence features, were found to be associated with T2D, while MALAT1 expression and KCNQ1OT1, ANRIL, and MEG3 sequence features were found to be associated in the second cohort. MEG3 was identified by SHAP as the dominant lncRNA in both cohorts. ML results were consistent with established statistical methods while additionally providing population- and subject-level disease association profiles linked to specific molecular feature types. The proposed framework advances mechanistic understanding of T2D and supports lncRNA-based precision medicine.

\end{abstract}


\maketitle

\section{Introduction}

Type 2 diabetes mellitus (T2D) is a chronic metabolic disorder 
characterized by progressive pancreatic $\beta$-cell dysfunction and 
peripheral insulin resistance, representing one of the most significant 
and rapidly expanding public health challenges of the twenty-first 
century~\cite{galicia-garcia_pathophysiology_2020, sun_idf_2022}. 
Despite decades of research into the protein-coding genome, the molecular mechanisms
underlying $\beta$-cell failure in T2D 
remain incompletely understood, and therapeutic options targeting the 
root causes of disease progression remain limited~\cite{halban_cell_2014}. Increasing 
evidence suggests that a substantial proportion of the regulatory 
landscape governing $\beta$-cell identity, function, and stress response 
operates at the level of non-coding RNA, particularly lncRNAs~\cite{wilson_role_2021}.

LncRNAs are transcripts exceeding 200 nucleotides in length that lack 
significant protein-coding potential yet exert diverse and often 
indispensable regulatory roles across virtually all biological 
processes. These include chromatin remodeling, transcriptional 
regulation, post-transcriptional mRNA stability, and the organization 
of nuclear architecture. Unlike microRNAs, lncRNAs execute their 
functions through a combination of sequence-specific interactions and 
secondary structure-dependent mechanisms, making their regulatory 
repertoire considerably more complex and context-dependent~\cite{statello_gene_2021}. Their sequence and structural-conformations control how they interact with proteins, DNAs, or other RNAs. These functional aspects are not captured by expression data alone, as highlighted in ~\cite{rinn_genome_2012, mercer_structure_2013}. In 
the context of T2D, several lncRNAs have been implicated in the 
regulation of insulin secretion, $\beta$-cell proliferation, and 
apoptotic pathways. Among these, 10 lncRNAs:
\textit{MALAT1}, \textit{MEG3}, \textit{MIAT}, \textit{CDKN2B-AS1} (ANRIL), \textit{GAS5}, \textit{KCNQ1OT1}, \textit{H19},  \textit{BCYRN1}, \textit{XIST}, and \textit{HOTAIR} have been reported as 
differentially expressed or functionally relevant in T2D and associated metabolic phenotypes~\cite{pandey_current_2022, sathishkumar_linking_2018, akella_compendium_2025, he_lncrnas_2017}. 

However, the majority of existing studies rely on conventional differential expression 
analysis of gene expression values or single-modality approaches, which neither integrates the full 
molecular profile of lncRNAs nor provides subject-level resolution of 
disease association. 
Moreover, gene expression is highly context-dependent, varying across tissue type, physiological condition, and disease stage, so it may reflect consequences rather than causes of T2D, as shown in ~\cite{iyer_landscape_2015}.
Because T2D is a complex disease involving multiple regulatory layers, relying on a single omics view gives an incomplete picture. Numerous studies have highlighted that integrative multi-omics approaches offer improved insights into disease mechanisms and biomarker discovery by capturing complementary information across different biological layers ~\cite{hasin_multi-omics_2017, karczewski_integrative_2018}. However, such approaches are often associated with high experimental cost, increased computational complexity, and limited data availability across cohorts. While classical statistical measures such as effect size of Cliff's $d$ ~\cite{cliff_dominance_1993} and Log$_2$ Fold Change 
(Log$_2$FC) ~\cite{love_moderated_2014} values are widely used to identify differentially expressed lncRNAs at the 
population level, they cannot generalize to unseen subjects.

To address these gaps, we present a comprehensive machine learning (ML) 
framework for lncRNA-disease association prediction that integrates 
sequence-, structure-, and expression-derived features 
extracted exclusively from RNA-seq data. This unified approach reduces dependency on additional omics platforms while still enabling comprehensive characterization of biological systems. The framework operates at two 
levels: (i) a population-specific lncRNA-T2D association classification framework that evaluates the 
discriminative capacity of each selected lncRNA independently in specific population, and find candidate lncRNAs (i.e., those lncRNAs out of 10 reported lncRNAs, which exhibit better association with T2D based on their better ML model predictive performance) for different combination of feature sets as described in coming sections, 
(ii) an integrated ML framework using only the selected features of lncRNAs which are found to be associated in (i), i.e., from candidate lncRNAs only. It not only provides T2D risk predictions in individuals but also provides the subject-wise association estimates using SHAP (SHapley Additive exPlanations) analysis \cite{ponce-bobadilla_practical_2024}, enabling quantification of individual-level lncRNA 
contribution to disease classification, a capability absent in 
existing approaches. 
Although, demonstrated here for T2D using pancreatic 
islet RNA-seq data from well-characterized European cohorts, the 
framework is designed to be fully generalizable to other diseases, 
tissues, and population cohorts, with population-specific retraining 
supported by design.

In this study, we apply the framework to ten lncRNAs previously 
implicated in T2D mentioned above, using pancreatic islet 
RNA-seq data from two independent publicly available cohorts. The 
primary objective is to validate the framework on previous literature validated T2D-associated 
lncRNAs as a proof-of-concept, establishing the methodological foundation 
for subsequent discovery of novel lncRNA-disease associations in 
population-specific cohorts. To our knowledge, this represents the first 
framework to integrate sequence composition, secondary structure, and 
RNA-seq expression features within a unified, machine 
learning framework for lncRNA-disease association prediction at both 
cohort and subject resolution, along with capability to infer the feature type (due to Expression, Structure, etc.) association.
The main contributions of this work are summarized as follows:
\begin{itemize}
	\item An end-to-end bioinformatics-ML framework is proposed for 
	population- and subject-specific lncRNA-T2D association 
	identification, extracting expression, structure, and sequence 
	features solely from RNA-seq data.
	
	\item Population-specific lncRNA-T2D associations are identified, 
	with results consistent with established statistical methods, 
	supporting lncRNA-based precision medicine and improved 
	mechanistic understanding of T2D.
	
	\item SHAP analysis enables cohort-level dominant lncRNA identification 
	and per-subject feature contribution profiling for precision 
	medicine interpretation.
\end{itemize}

The structure of this paper is as follows. Section~\ref{sec:materials} describes 
the datasets, feature extraction methodology, and ML 
framework design. Section~\ref{sec:Results} presents the classification results. Section~\ref{sec:conclusion}
discusses the conclusion of this study, framework limitations, and 
directions for future work. 

\section{Materials and Methods}\label{sec:materials}

\subsection{Data Sources and Dataset Description}

In this study, publicly available RNA-sequencing datasets related to Type-2 Diabetes (T2D) were obtained from the NCBI Gene Expression Omnibus (GEO) repository to investigate patterns associated with T2D. Two Bulk RNA-sequencing datasets, GSE159984 ~\cite{marselli_persistent_2020} and GSE164416 ~\cite{wigger_multi-omics_2021}, containing gene expression profiles along with raw sequencing reads obtained from human pancreatic islets from individuals with T2D and non-diabetic controls (NDC).

The dataset GSE159984 contains bulk RNA-sequencing profiles of human pancreatic islet cells obtained from organ donors (Table \ref{tab:datasets}). The original dataset consists of 58 control subjects and 28 T2D subjects. In the present study, 46 control subjects and 20 T2D subjects were retained for downstream analysis. These were generated using bulk RNA-sequencing technology with a single-end library layout. 

The dataset GSE164416 contains bulk RNA-sequencing data derived from pancreatic islet subjects obtained from patients undergoing pancreatectomy surgery. These subjects belong to the Pancreatectomized Patient Program (PPP) cohort, which includes T2D and NDC cases. The selected dataset consists of 39 T2D subjects and 18 NDC subjects and was generated using high-throughput RNA-sequencing technologies.

Both datasets contain gene expression measurements and raw sequencing reads for thousands of genes across multiple biological subjects, enabling computational analysis for disease classification and biomarker discovery. The expression datasets were downloaded directly from the GEO repository and raw sequence reads were downloaded using the SRA toolkit (SRA number on GEO repository) and used for subsequent preprocessing, feature extraction and ML analysis.

The datasets correspond to independent donor cohorts from distinct European study populations, and were therefore processed separately to preserve cohort-specific characteristics and integrating the datasets or assessing cross-dataset performance may lead to suboptimal results due to batch effects and inherent population-specific variations.

This study used publicly available GEO datasets; therefore, no additional ethical approval or patient consent was required.

\begin{table}
	\caption{Summary of the gene expression datasets used in this study.}
	\label{tab:datasets}
	\begin{tabular}{c c c c}
		Dataset &  T2D & NDC & Sequencing Type \\
		\midrule
		{\color{blue}\underline{\href{https://www.ncbi.nlm.nih.gov/geo/query/acc.cgi?acc=GSE159984}{GSE159984}}}  & 20 & 46 & RNA-seq (single-end; removed paired-end) \\
		{\color{blue}\underline{\href{https://www.ncbi.nlm.nih.gov/geo/query/acc.cgi?acc=GSE164416}{GSE164416}}} & 39 & 18 & RNA-seq (single-end) \\
	\end{tabular}
\end{table}

\subsection{Data Preprocessing}

\begin{enumerate}[label=(\roman*)]

	\item \textbf{Expression dataset preprocessing}
	
	The gene expression matrices were downloaded from the GEO database and processed prior to ML analysis. Preprocessing tasks included adding gene name column corresponding to the given ncbi id column, transpose of the matrix, adding Category column for matching subject number using metadata file downloaded from GEO repository, and filtering of subjects with single library-layout to avoid batch effects within the same dataset (GSE159984). In GSE164416, we retained only T2D and NDC categories as per our study requirements.

	\item \textbf{RNA Sequence Processing and Subject-Specific lncRNA Feature Extraction}
	
	Raw sequencing reads present at the GEO datasets, were downloaded using the SRA toolkit, in the \textit{.sra} file fromat. For further processing using bioinformatics tools, they were converted to {.fastqz} format. 
	
	A comprehensive RNA-seq processing framework was implemented to derive subject-specific lncRNA sequences variants information. Raw sequencing reads were first subjected to quality assessment using FastQC ~\cite{simon_fastqc_2010}, followed by adapter and low-quality base trimming using Trim Galore, which is an open-source quality-control wrapper ~\cite{felix_trim_2012}. This step improves the quality of reads and enhances downstream analysis reliability. The processed reads were then aligned to the human reference genome (hg38) v45 using STAR aligner in two-pass mode to improve splice junction detection. Post-alignment processing included duplicate marking, addition of read group information, and splitting of reads spanning introns to ensure compatibility with downstream variant calling workflows.
	
	Variant calling was performed using the GATK HaplotypeCaller in GVCF mode, restricted to annotated lncRNA regions ~\cite{mckenna_genome_2010}. A custom BED file defining the genomic coordinates of the selected 10 lncRNAs was created to constrain the analysis to regions of interest. Joint genotyping and variant filtration were subsequently applied using recommended RNA-seq hard-filtering criteria. To minimize false positives arising from RNA editing events, common A$\rightarrow$G and T$\rightarrow$C substitutions were excluded. High-confidence variants were then extracted and summarized for each subject.
	
	To construct subject-specific lncRNA sequences, identified variants were applied to the reference genome using a consensus-based approach, generating personalized lncRNA FASTA sequences for each subject. In parallel, transcript-level expression quantification was performed using featurCounts, followed by normalization to Transcripts Per Million (TPM) values. This unified framework enabled the extraction of expression, sequence, and structural information from RNA-seq data, facilitating multi-level characterization of lncRNAs without requiring additional omics datasets.
	
\end{enumerate}

\subsection{Feature Extraction}
\label{sec:features}

A dedicated feature extraction framework was developed to generate 
comprehensive and biologically meaningful representations of the ten 
lncRNAs. Rather than relying on a single feature type, the framework 
integrated multiple complementary modalities spanning sequence composition, 
secondary structure, and transcriptional expression. Together, these modalities 
provided a unified and flexible input space for downstream ML 
models and enabled systematic assessment of the relative contribution of 
each feature category to lncRNA-disease association prediction. The 
complete feature matrix comprised 90,501 features per lncRNA prior to 
downstream selection. Features were adapated from the established resources including PyFeat toolkit ~\cite{muhammod_pyfeat_2019, bonidia_feature_2021}, published ML based frameworks and statistical descriptors. Where existing implementations required adaptation, custom Python scripts were developed.

\begin{enumerate}[label=(\roman*)]

	\item \textbf{Sequence-Derived Features -} Sequence-level features capture the 
	compositional, statistical, and signal-based properties of lncRNA primary 
	sequences, collectively encoding both local sequence motifs and global 
	sequence organization. The following descriptors were extracted using the 
	PyFeat toolkit~\cite{muhammod_pyfeat_2019} and custom implementations 
	where required.

	\begin{enumerate}[label=(\alph*)]

		\item $k$-mer frequency profiles were computed for $k = 3, 4, 5, 6, 7, 8$ 
		over the alphabet $\Sigma = \{A, C, G, T\}$, enumerating the absolute 
		occurrence of all $4^{k}$ possible nucleotide subsequences per sequence. 
		Eight classes of $k$-gap descriptors were extracted at gap $g = 1$, 
		covering all combinations of mono-, di-, and trinucleotide units at both 
		anchor and gap positions ~\cite{muhammod_pyfeat_2019}. 
		\item Pseudo-dinucleotide composition was computed following the PseKNC 
		formulation~\cite{chen_pseudo_2015} with sequence-order weighting parameters 
		$\lambda = 2$ and $w = 0.05$, incorporating both nearest-neighbor 
		physicochemical correlations and sequence-order effects beyond simple 
		composition. \item Z-curve descriptors~\cite{zhang_z_1994, muhammod_pyfeat_2019} 
		decomposed each sequence into three periodic components $x$, $y$, 
		and $z$ curves reflecting purine/pyrimidine, amino/keto, and 
		weak/strong hydrogen bond distributions respectively, providing a 
		geometric representation of nucleotide organization.
		\item GC content, GC 
		count, AT/GC ratio, GC skew, and AT skew were computed as global 
		compositional and strand-asymmetry indices~\cite{muhammod_pyfeat_2019}. 
		\item Shannon entropy was computed with $k = 10$ to assess the informational 
		complexity of each sequence~\cite{muhammod_pyfeat_2019}. 
		\item Fourier 
		transform-based spectral features were extracted after aplying $Z\_curve$ numerical mapping at resolution $r = 2$, 
		converting the nucleotide sequence into a numerical representation and 
		applying the discrete Fourier transform to capture periodic signals 
		embedded in the primary sequence~\cite{muhammod_pyfeat_2019}. 
		\item Complex 
		network-based descriptors were derived by representing the lncRNA 
		sequence as a graph with $k = 3$ and threshold $t = 10$, from which 
		topological properties were computed to capture higher-order sequence 
		dependencies not accessible to linear 
		descriptors~\cite{muhammod_pyfeat_2019}.
	\end{enumerate}
	
	\item \textbf{Structure-Derived Features -}Secondary structure features reflecting the functional relevance of lncRNA folding were extracted using the Linearfold package, including minimum free energy (MFE), base-pairing profiles, and proportions of paired and unpaired bases~\cite{huang_linearfold_2020}.
	
	\item \textbf{Expression-Derived Features -} 
	\begin{enumerate}[label=(\alph*)]
		\item TPM values were downloaded directly from publicly 
		available GEO datasets SRA Run selector ~\cite{marselli_persistent_2020, wigger_multi-omics_2021}  and were used without additional preprocessing. TPM provides a measure of expression level comparable across subjects. 
		\item Log$_2$ Fold-Change (log$_2$FC) values were computed using the formula: \\
		$\log_2 FC_i = \log_2 \left( \frac{TPM_i}{\mathrm{average}(TPM_{\text{NDC}})} \right)$ \\
		i.e., by taking the log$_2$ of ratio of each subject's TPM value with average TPM value of controls, where i denotes subject. This feature quantifies the magnitude and directionality of differential regulation per lncRNA.
	\end{enumerate}
	
\end{enumerate}

After feature extraction, the total number of sequence-derived features was 90{,}494. \texttt{TPM} and \texttt{Log$_2$FC} were expression features. \texttt{MFE}, \texttt{paired\_bases}, \texttt{unpaired\_bases}, \texttt{pair\_ratio}, and \texttt{norm\_MFE} were structural features.

\subsection{Feature Selection}
\label{sec:feature_selection}
Feature selection is required in order for ML models to work properly, as the large number of features are overwhelming to them and uninformative features leads to the artifacts in the predicitons. Using feature selection, we filtered out only the most informative features and provided them to the model to improve prediction performance. In this study, a feature-selection cap or maximum number of features were set to 15 features given the small number of subject size in the datasets used for this study, using the formula: $max\_features$ = one-fifth of the training subject size as it is helpful in preventing overfitting. 

Feature selection proceeded in three sequential stages: (i) a variance 
threshold filter to remove low-variance features, (ii) ANOVA F-test with k=100 to 
retain best 100 features exhibiting statistically significant group 
differences, and (iii) ExtraTrees-based feature importance ranking was used to identify the most discriminative features~\cite{geurts_extremely_2006}. The model was configured with controlled tree depth ($\text{max\_depth}=4$) and minimum subject constraints ($\text{min\_samples\_split}=4$, $\text{min\_samples\_leaf}=3$) to reduce overfitting on the limited dataset. An ensemble size ($n_{\text{estimators}}=100$) ensures stable importance estimates, $\text{max\_features}=\sqrt{d}$, where $d$ = total number of input features, promotes feature diversity across trees. To address imbalance $\text{class\_weight}=\text{balanced}$ \cite{taghizadeh_breast_2022}.

\subsection{Machine Learning Framework}
\label{sec:mlframework}

\begin{figure}[tb]
	\centering
		\includegraphics[width=3.5in]{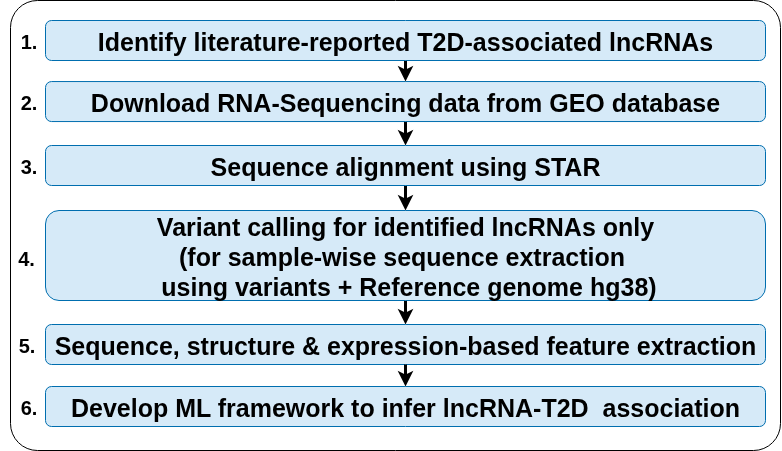}
	\caption{Overall workflow of the proposed framework}
	\label{fig:methodology}
\end{figure}

\begin{figure*}[t]
	\centering
		\includegraphics[width=7.16in]{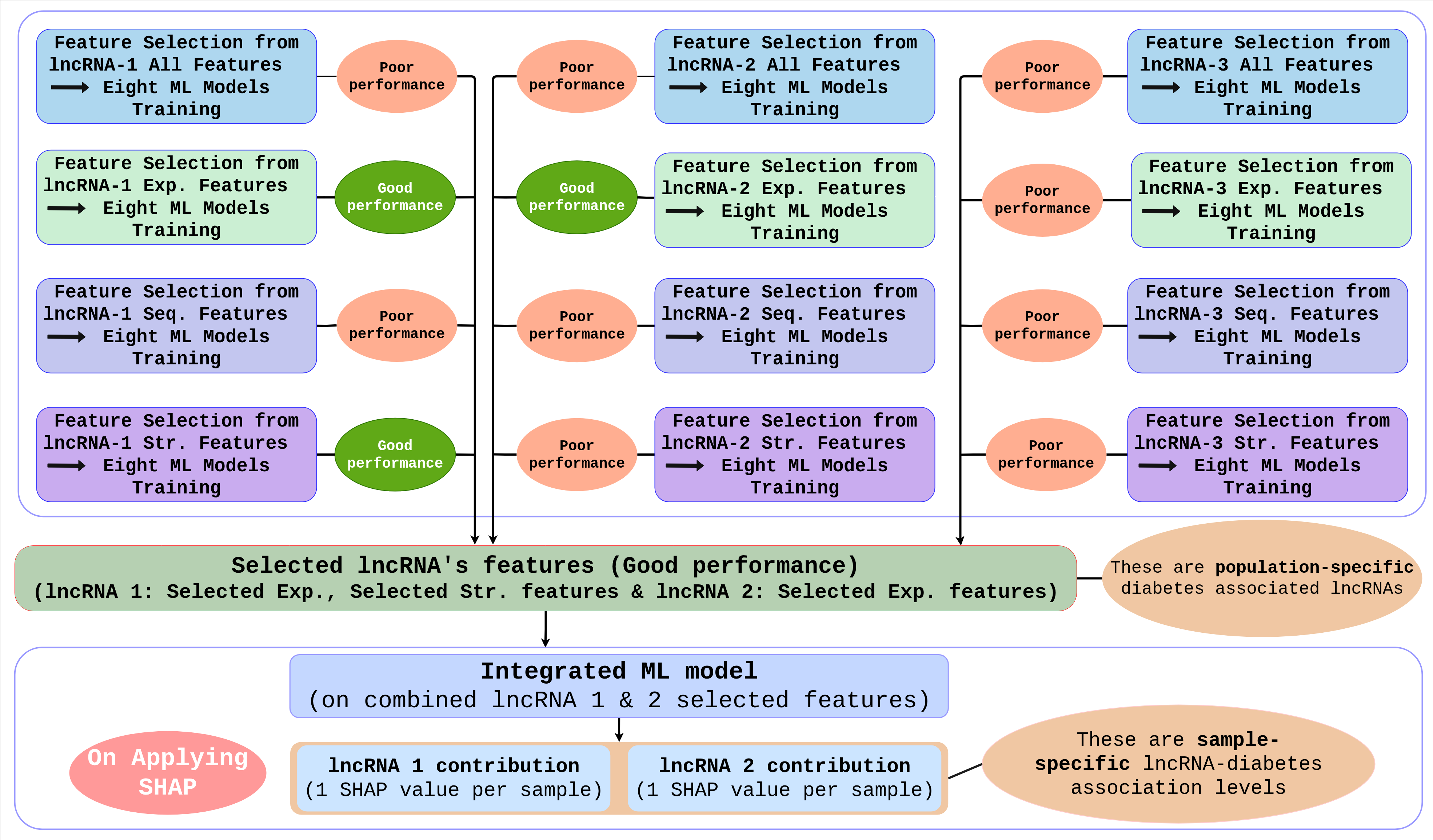}
	\caption{Overview of the proposed lncRNA–T2D association framework (Exp.: Expression; Seq.: Sequence; Str.: Structure).}
	\label{fig:framework}
\end{figure*}

The overall workflow of the proposed framework, including data flow from RNA-seq input through 
feature extraction to association prediction, is summarized in 
Figure~\ref{fig:methodology}.
An overview of the proposed ML modeling framework is illustrated 
in Figure~\ref{fig:framework}. For each candidate lncRNA, the integrated 
feature matrix described in Section~\ref{sec:features} serves as input 
to two classification frameworks, an individual lncRNA feature-set based ML framework 
evaluating each lncRNA independently, and an integrated ML framework trained on associated lncRNAs features only, to provide per-subject association levels.

Both frameworks employ the following eight ML classifiers:
Random Forest (RF), an ensemble method aggregating multiple 
decision trees via majority voting to reduce overfitting on high-dimensional biological data ~\cite{breiman_random_2001}; 
Extra Trees (ET), which introduces additional randomness in feature selection and split-point 
determination, yielding lower variance and suiting small-sample settings ~\cite{geurts_extremely_2006}; 
Logistic Regression (LR), a linear probabilistic classifier modelling the log-odds of class membership and serving as an interpretable baseline with well-calibrated probability estimates \cite{hosmer_applied_2000}; 
K-Nearest Neighbors 
(KNN), a non-parametric instance-based classifier assigning labels by majority vote among the 
$k$ closest training subjects in feature space \cite{zhang_introduction_2016};  
Support Vector Classifier (SVC) is a supervised learning algorithm based on the principles of Support Vector Machine. It works by finding the optimal hyperplane that maximizes the margin between different classes in the feature space. For non-linearly separable data, it uses kernel functions (e.g., linear, polynomial, RBF) to map data into higher-dimensional spaces where a separating boundary can be found \cite{cortes_support-vector_1995}. Linear 
Discriminant Analysis (LDA), a generative classifier projecting data onto a subspace 
maximizing between-class separation under Gaussian class-conditional assumptions \cite{tharwat_linear_2017}; Decision 
Tree (DT), a non-parametric model recursively partitioning the feature space by minimizing 
impurity, offering high interpretability \cite{quinlan_induction_1986}; and Naïve Bayes (NB), a probabilistic classifier 
grounded in Bayes' theorem that assumes conditional feature independence, enabling efficient 
inference under limited subject sizes \cite{peretz_naive_2024}.


\subsection{Model Evaluation Strategy}
\label{sec:evaluation}

\begin{table*}[h]
	\centering
	\caption{Performance-based classification criteria (thresholds apply to K-fold, LOOCV, and hold-out metrics).}
	
	\label{tab:thresholds}
	\begin{tabular*}{\textwidth}{l@{\extracolsep{\fill}}ccccc}
		\hline
		\textbf{Category} & \textbf{AUC} & \textbf{MCC} & \textbf{Recall} & \textbf{Specificity} & \textbf{F1-score} \\
		\hline
		Good   & $\ge 0.80$ & $\ge 0.50$ & $\ge 0.70$ & $\ge 0.70$ & $\ge 0.70$ \\
		Moderate & $\ge 0.70$ & $\ge 0.30$ & $\ge 0.60$ & $\ge 0.60$ & $\ge 0.60$ \\
		Weak     & $\ge 0.60$ & $\ge 0.20$ & $\ge 0.50$ & $\ge 0.50$ & $\ge 0.50$ \\
		Failed   & \multicolumn{5}{c}{Does not satisfy the above criteria} \\
		\hline
	\end{tabular*}
\end{table*}

The predictive performance of the ML models was evaluated using three evaluation techniques that include: Stratified K-Fold CV~\cite{fushiki_estimation_2011}, Leave-One-Out Cross-Validation
(LOOCV)~\cite{cheng_efficient_2017}, and Repeated Hold-Out evaluation~\cite{raschka_model_2018}. They provide a complementary and independent assessment of generalization. In Stratified k-Fold, full dataset gets splited in k number of mutually exclusive subsets (folds) such that class distribution is preserved (i.e., preserving the same ratio of T2D and NDC subjects in both training and testing sets). During evaluation, k iterations are performed; in each iteration, one fold is used as the test set while the remaining k-1 folds are used for training. This process ensures that every subject is used for both training and validation exactly once, while preserving class proportions across all folds.
LOOCV is a special case of k-fold cross-validation where the number of folds equals the number of subjects ($k = N$), each subject is iteratively used once as a test instance while 
the remaining N-1 subjects are used for training. This process is repeated N times such that each subject is used exactly once as a test instance. The final performance is obtained by averaging the results across all iterations.
In repeated hold-out validation, dataset is randomly split into training and test sets multiple times. In each iteration, a fixed proportion of the data (e.g., $80\%$) is used for training and the remaining portion for testing. Model is trained and evaluated across multiple such random splits, and final performance is obtained by averaging the results.

Following performance metrics were computed across all evaluation strategies: 
Area Under the ROC Curve (AUC), Matthews Correlation Coefficient (MCC)~\cite{chicco_advantages_2020}, F1-score, accuracy,  recall, and specificity.

Performance was additionally classified into qualitative tiers as \textit{Failed}, \textit{Weak}, \textit{Moderate}; and \textit{Good} based upon selection criteria as described in the Table~\ref{tab:thresholds}. 

Thresholds were chosen to reflect the modest discriminative capacity expected under small subject sizes and
class imbalance, rather than the stricter standards applicable to balanced, large-cohort classifiers \cite{sujon_accuracy_2025}.  

\subsection{Preprocessing and Statistical Evaluation Metrics}
\label{sec:preprocessing}

This subsection describes the preprocessing steps and statistical measures employed to ensure robust and unbiased model development and evaluation.

\begin{enumerate}[label=(\roman*)]

	\item \noindent \textbf{Standard Scaling}
	To normalize feature distributions, standardization was applied using the z-score transformation:
	\begin{equation}
		z = \frac{x - \mu}{\sigma}
	\end{equation}
	where $x$ is the original feature value, $\mu$ is the mean, and $\sigma$ is the standard deviation of the feature. This ensures that all features contribute equally to model training.
	
	\item \noindent \textbf{Effect Size Measurement}
	
	Cliff’s $d$~\cite{cliff_dominance_1993} was used to quantify the effect size between the T2D and NDC groups at the feature level in a non-parametric manner. 
	
	For a feature $f$, let $X_1^{(f)}=\{x_{1,1}^{(f)},...,x_{1,n_1}^{(f)}\}$ and $X_2^{(f)}=\{x_{2,1}^{(f)},...,x_{2,n_2}^{(f)}\}$ denote the feature values across all T2D and NDC subjects, respectively, where $x_{k,i}^{(f)}$ represents the value of feature $f$ for the $i$-th lncRNA in group $k$.

	Cliff’s $d$ value for feature $f$ is computed as:
	\begin{equation}
		d_f = \frac{1}{n_1 n_2} \sum_{i=1}^{n_1} \sum_{j=1}^{n_2} 
		\text{sgn}\big(x_{1,i}^{(f)} - x_{2,j}^{(f)}\big)
	\end{equation}
	where $\text{sgn}(\cdot)$ returns $+1$ if the feature values in T2D are larger than NDC values, $-1$ if smaller, and $0$ if equal. Thus, $d_f$ measures how often values from the T2D group exceed those from the NDC group across all pairwise comparisons. $d_f$ yields a bounded statistic $d_f \in [-1, 1]$.
	
	This formulation does not assume normality or equal variances and is therefore well-suited for high-dimensional and potentially skewed biological data. Effect size magnitudes were interpreted following Romano et al.~\cite{romano_appropriate_2006}.
	
	\noindent\textit{Application to expression features.} For each lncRNA, a single 
	Cliff's $d$ value was computed between the normalised expression levels of the 
	T2D group ($n_{\mathrm{T2D}}$ subjects) and the NDC group ($n_{\mathrm{NDC}}$ 
	subjects) in each cohort independently. This scalar 
	effect size characterises the direction and magnitude of differential expression 
	at the lncRNA level and complements the $\log_2$ fold-change by remaining robust 
	to outliers.
	
	\noindent\textit{Application to sequence-derived features.} For each lncRNA, a 
	feature matrix of dimension $N_s \times 90{,}494$ was constructed, where $N_s$ 
	denotes the number of subjects and the 90,494 columns. Cliff's $d$ was computed independently for each feature $i$ between 
	the T2D and NDC groups:
	\begin{equation}
		\begin{split}
			d_i &= \frac{1}{n_{\mathrm{T2D}} n_{\mathrm{NDC}}} 
			\sum_{p=1}^{n_{\mathrm{T2D}}} \sum_{q=1}^{n_{\mathrm{NDC}}} 
			\mathrm{sgn}\!\left(f_{i,p}^{\mathrm{T2D}} - f_{i,q}^{\mathrm{NDC}}\right),\\
			&\quad i = 1,\ldots,90{,}494
		\end{split}
	\end{equation}
	where $f_{i,p}^{\mathrm{T2D}}$ and $f_{i,q}^{\mathrm{NDC}}$ denote the value of 
	feature $i$ for T2D subject $p$ and NDC subject $q$, respectively. Two summary 
	statistics were derived from the resulting distribution of $\{d_i\}$:
	
	\begin{enumerate}[label=(\roman*)]
		\item \textit{Peak discriminability} - the maximum absolute \textbf{Cliff’s Delta ($d$)} value across all sequence features, representing the Cliff’s $d$ reported in this study.
		\begin{equation}
			d_{\mathrm{seq}}^{\max} = \max_{i \in \{1,..,90494\}} |d_i|
		\end{equation}
		This identifies whether any single sequence feature achieves strong 
		separation between T2D and NDC subjects for a given lncRNA.
		
		\item \textit{Breadth of signal} - the count of features exceeding the 
		medium effect size threshold:
		\begin{equation}
			N_{\mathrm{mod}} = \sum_{i=1}^{90{,}494} \mathbf{1}(|d_i| > 0.33)
		\end{equation}
		where $\mathbf{1}(\cdot)$ is the indicator function. A large $N_{\mathrm{mod}}$ 
		indicates that the discriminative signal is distributed broadly across the 
		sequence representation, whereas a small $N_{\mathrm{mod}}$ with a high 
		$d_{\mathrm{seq}}^{\max}$ indicates that the signal is concentrated in a 
		narrow subset of features.
	\end{enumerate}
	
	Together, $d_{\mathrm{seq}}^{\max}$ and $N_{\mathrm{mod}}$ provide a 
	complementary characterisation of each lncRNA sequence-level discriminability: 
	the former captures peak separation and the latter captures the extent to which 
	the T2D/NDC difference is reflected across the full sequence feature space. These 
	statistics were computed separately for each cohort and reported alongside the 
	expression-level Cliff's $d$ in the per-lncRNA characterisation Table~\ref{tab:discriminability_both}.
	
	\item \noindent \textbf{SHAP values}
	
	SHAP (SHapley Additive exPlanations) \cite{ponce-bobadilla_practical_2024} values are grounded in Shapley values from cooperative game theory and attribute a model’s prediction to individual features. For a given instance $x$ and feature $i$, the SHAP value $\phi_i(x)$ is defined as:
	
	\begin{equation}
		\phi_i(x) =
		\sum_{S \subseteq F \setminus \{i\}} \frac{|S|!\,(|F| - |S| - 1)!}{|F|!}
		\left[
		f_{x}(S \cup \{i\}) - f_{x}(S)
		\right],
	\end{equation}
	
	where $F$ is the set of all features, $S$ is a subset of features excluding $i$, and $|F|$ denotes the total number of features. The symbol $!$ denotes the factorial. The term $f_{x}(S)$ represents the expected model output conditioned on the features in $S$ taking their values from instance $x$.
	
	The weighting term ensures that each subset $S$ contributes proportionally, yielding the average marginal contribution of feature $i$ across all possible subsets.
	
\end{enumerate}

\subsection{Population-Specific lncRNA-T2D Association Classification Framework}
\label{sec:perlncrna}

\begin{algorithm*}[tb]
	\small

	\caption{Population-Specific lncRNA-T2D Association Classification Framework}
		\label{alg:algorithm1}
	\begin{algorithmic}[1]

		\Require lncRNA feature file; $K\_fold{=}5$; $R{=}100$ repeats
		\Ensure Aggregated K-Fold, LOOCV and Hold-out metrics summary including performance tier;

		\State $\mathcal{L} \leftarrow \{\text{MALAT1}, \text{MEG3}, \text{MIAT}, \text{ANRIL}, \text{GAS5}, \text{KCNQ1OT1}, \text{H19}, \text{BCYRN1}, \text{XIST}, \text{HOTAIR}\}$
		
		\For{each lncRNA $\ell \in \mathcal{L}$}
		\For{each feature-set $f \in \{\text{Expression}, \text{Sequence}, \text{Structure}, \text{All}\}$}
		
		\State \textbf{// Preprocessing}
		\State Load feature-set $\mathcal{F}$ data $(\mathbf{X}, \mathbf{y})$ corresponding to lncRNA $\mathcal{L}$; map labels $\{\text{T2D}\!\rightarrow\!1,\ \text{ND}\!\rightarrow\!0\}$
		
		\State $\mathbf{X} \leftarrow \mathbf{X}_{\textsc{SelectFeatures}(\mathbf{X}, \mathbf{y})}$; // In 3 steps: VarianceThreshold $\rightarrow$ ANOVA $\rightarrow$ ExtraTrees Classifier
		
		\For{each model $m \in \{\text{RF, ET, LR, LDA, KNN, NB, DT, SVC}\}$}

		\State \textbf{// K-Fold CV }
		\State Partition $(\mathbf{X}, \mathbf{y})$ into $K{=}3$ folds
		\For{each fold}
		\State Split: $(\mathbf{X}_{\text{tr}}, \mathbf{y}_{\text{tr}})$ train;\; $(\mathbf{X}_{\text{te}}, \mathbf{y}_{\text{te}})$ test
		\State Apply \textsc{StandardScaler}: fit\_transform on $\mathbf{X}_{\text{tr}}$, transform on $\mathbf{X}_{\text{te}}$
		
		\State Train model $m$ on $\mathbf{X}_{\text{tr}}$; Predict class labels $\hat{y}$ (\textit{y\_pred}) and probabilities $\hat{p}$ (\textit{y\_prob}) on $\mathbf{X}_{\text{te}}$
		\EndFor
		\State Aggregate fold predictions; compute: AUC, Accuracy, MCC, Recall, Specificity, F1

		\State \textbf{// LOOCV }
		\For{each subject $i = 1,\ldots,n$}
		\State Split: $(\mathbf{X}_{-i},\mathbf{y}_{-i})$ train;\; $(\mathbf{x}_i,y_i)$ test
		
		\State Apply \textsc{StandardScaler}: fit\_transform on ${\mathbf{X}}_{-i}$ and transform on ${\mathbf{x}}_i$
		
		\State Train model $m$ on $\mathbf{X}_{-i}$ and Predict $\hat{y}_i, \hat{p}_i$ on ${\mathbf{x}}_i$
		\EndFor
		\State Aggregate: AUC, ACC, MCC, F1, Recall, Specificity

		\State \textbf{// Repeated Hold-out }
		\For{$r = 1,\ldots,100$}
		\State $(\mathbf{X}_{\text{tr}},\mathbf{X}_{\text{te}}) \leftarrow \textsc{StratifiedSplit}(80\%/20\%)$
		\State Apply \textsc{StandardScaler}: fit\_transform on $\mathbf{X}_{\text{tr}}$, transform on $\mathbf{X}_{\text{te}}$
		\State Train model $m$, evaluate on test
		\EndFor
		\State Report mean $\pm$ SD (AUC, ACC, MCC, F1, Recall, Specificity)

		\State \textbf{// Final performance classification}
		\State Assign \{Good, Moderate, Weak, Failed\} based on K-Fold, LOOCV, and Hold-out metrics
		\EndFor \EndFor \EndFor
		
	\end{algorithmic}
\end{algorithm*}

\begin{figure*}[tb]
	\centering
		\includegraphics[width=7.16in]{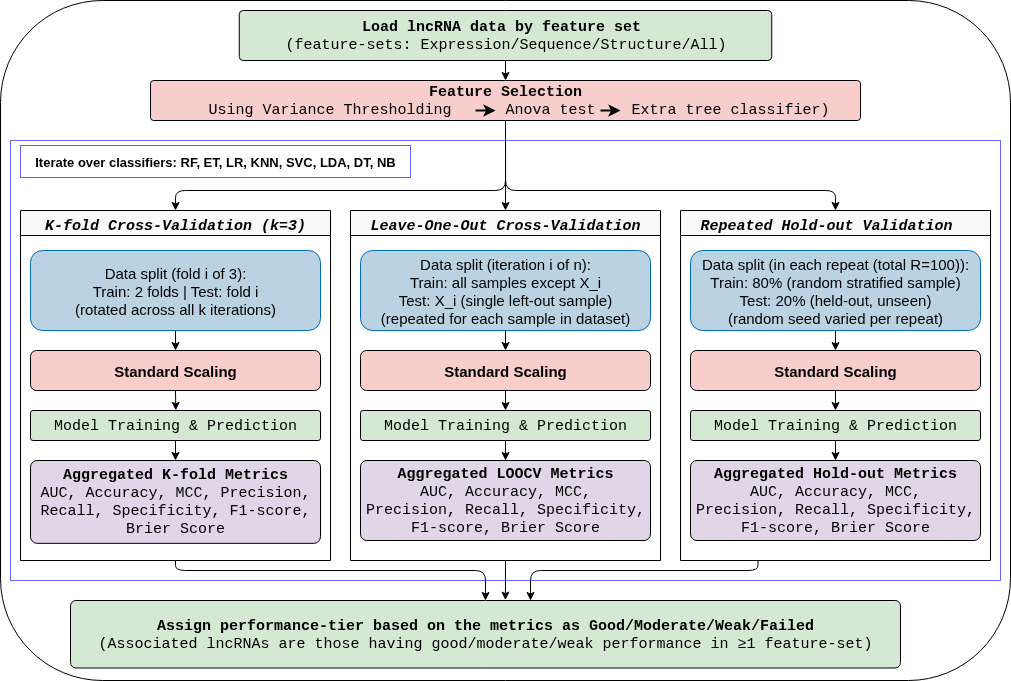}
	\caption{Detailed flow of the proposed ML framework for finding
		population-specific lncRNA-disease associations.}
	\label{fig:ml_pipeline_obj1}
\end{figure*}

Our hypothesis says that, if an lncRNA have discriminative properties to classify T2D vs NDC based upon its features in a population, then it suggests there exists some association of that lncRNA in that specific population and this discriminativeness can be confirmed if machine learning models are able to predict well on these features. To further extend this, if an lncRNAs feature-set configuration is independently able to provide better ML model performance, it suggests that particular lncRNA have association with T2D for those feature-set properties.
To evaluate the discriminative capacity of each candidate lncRNA independently, a single-lncRNA based classification framework was developed, which returns the performance of eight ML classifiers trained separately on each lncRNA feature-set combination. For each lncRNA, separate feature matrices were constructed for each feature-set, as described in Section~\ref{sec:features}.

In this framework, we first defined the lncRNAs name, feature-sets, name and parameters of eight ML classifiers. 
Then for each lncRNA, the following things were done:
Looped through all the feature-sets, in each loop, loaded current lncRNA dataset for current feature-set and did
data preprocessing such as labels mapping, removing duplicates, etc. 
Feature-selection was done once on the full dataset to identify the features with higher importance and only these features were kept while discarding the uninformative ones according to the feature selection procedure define in Section \ref{sec:feature_selection}. 
Once preprocessing and feature selection were done, for each ML classifier we iteratively did 
evaluation using three methods independently and their mean aggregated metrics were considered to assign the performance tier.
In K-fold CV, $k$ value was taken as $3$ due to small dataset, inside each iteration, standard scaling was performed such that $2$ training folds were used for fit and transform, and only transform one for the test fold. This ensured that there was no test data leakage during scaling among multiple iterations. After scaling, current ML model was trained and predictions were made for the test data and evaluation metrics were calculated. Once evaluation was done for all folds, mean aggregation was done on their metrics to find the final metrics of k-fold CV.
Similar to this, LOOCV evaluation was done by first performing the standard scaling, then model training, evaluation and finally aggreation by taking mean at the end to generate LOOCV metrics. In Repeated Hold-Out scheme, number of iterations were set at 100, training and testing size as $80\%$ anf $20\%$ respectively, then scaling, model training, evaluation and mean aggregation of metrics.
All three metrics were used to assign the performance tier based upon the thresholds mentioned in Table \ref{tab:thresholds}.
After this, next classifier was evaluated, and once all classifiers were evaluated, same procedure was followed for the next lncRNA. After all lncRNAs were evaluated, same procedure was adopted for the next feature-set. At the end, in results file we got $(\#\mathrm{feature\ sets} \times \#\mathrm{lncRNAs} \times \#\mathrm{classifiers}) 4 \times 10 \times 8 = 320$ rows, each containing $(\#\mathrm{metrics} \times \#\mathrm{evaluation\ schemes}) 8 \times 3 = 24$ metrics.

The lncRNAs with good, moderate or weak performance were considered to have population-specific associations, explaining their association strength according to their performance tier and also specifying the feature modality responsible for the association. These associated lncRNAs were named as candidate lncRNAs.

A detailed pseudo-code representation of the proposed framework is provided in the  Algorithm \ref{alg:algorithm1}. And detailed flowchart is provided in the Figure \ref{fig:ml_pipeline_obj1}.

\subsection{Subject-Specific lncRNA-T2D Association Classification Framework}
\label{sec:nested}

\begin{algorithm}[tb]
	\small

	\caption{Integrated Classification Framework for Subject-Specific lncRNA Associations}
		\label{alg:algorithm2}
	\begin{algorithmic}[1]

		\Require Results csv containing each lncRNA along with their selected features and performance tier.
		\Ensure SHAP attribution per lncRNA for each subject; performance tier of integrated model with metrics. 
		
		\Statex \textbf{// Build combiend feature-matrix and Preprocessing}
		\State Load the results CSV file and generate a dictionary mapping each lncRNA $\ell$ to its corresponding set of selected individual features (e.g., TPM, k-mer features), restricted to entries labeled as Good, Moderate, or Weak in performance.
		
		\For{each $\ell \in list$}
		\State Extract the selected features data from $\ell$ dataset and add these features to the combined feature matrix by prefixing each feature with the corresponding lncRNA name.
		\EndFor
		
		\State Follow the same steps as described in Algorithm~\ref{alg:algorithm1}, from preprocessing and feature selection up to the final step of performance tier assignment. // This time, no separate feature-set models are created, so only 8 entries will be present in the result, corresponding to the classifiers.
		
		\State \textbf{// Apply SHAP}
		\State Select the best-performing classifier based on evaluation metrics and compute per-subject SHAP values and aggregate them using sum to obtain lncRNA-wise contributions.
		\State Generate beeswarm, bar, and stacked plots for T2D cases to visualize lncRNA contributions.

	\end{algorithmic}
\end{algorithm}

To interpret subject-specific individual lncRNAs associations, SHAP was employed as part of a ML framework, consisting of combined lncRNAs features, which were found to be associated in single-lncRNA framework. This combination of candidate lncRNAs also provided improved discriminative performance relative to the individual lncRNAs.

For the multi-lncRNA framework, a dictionary mapping each lncRNA to its corresponding set of selected individual features was derived from the results of the section~\ref{sec:perlncrna}.  Only those lncRNA-feature combinations demonstrating either weak or moderate or good tier performance were retained in the dictionary.
For each of these candidate lncRNAs, their selected\_features were extracted and added to a combined feature-matrix consiting such features for all the candidate lncRNAs. 
Each selected feature column's name was prefixed with the corresponding lncRNA identifier (following the format $lncRNA\_feature$, e.g., $GAS5\_kmer\_ATGC$) to avoid namespace collision across lncRNAs in the combined feature matrix. All selected features were aggregated to form a combined feature matrix. This matrix was used for training the integrated ML classifier.

Rest of the process starting from feature selection, data preprocessing and evaluation to assigning the performance tier was same as stated in the section~\ref{sec:perlncrna}. And the result of this procedure was performance of integrated model on eight classifiers containing metrics for all three evaluation schemes.

Before applying SHAP, a final model having the best performance among the eight classifiers was selected using a hierarchical ranking strategy. 
\begin{equation}
	\begin{aligned}
		\text{Rank}(model) = (
		\text{Performance Tier}, \text{MCC}_{\text{HO}}, \text{AUC}_{\text{HO}}, \\
		\text{MCC}_{\text{LOOCV}}, \text{AUC}_{\text{LOOCV}}, \text{MCC}_{\text{KFold}}, \text{AUC}_{\text{KFold}})
	\end{aligned}
	\label{eq:model_score}
\end{equation}
Models labeled as \textit{Failed} were excluded, and the remaining candidates were sorted in descending order according to the strategy prioritising overall performance consistency, followed by Matthews correlation coefficient (MCC) and area under the ROC curve (AUC) across multiple validation protocols (hold-out, LOOCV, and k-fold). This approach emphasises robustness under class imbalance and stability over reliance on a single metric or validation scheme.

SHAP values were computed for each subject on this final model using TreeExplainer when applicable, with KernelExplainer used as a fallback.
SHAP values indicate how each feature increases or decreases a subject’s predicted T2D probability relative to the model’s average prediction.
Feature-level SHAP values were aggregated per lncRNA by taking sum contributions of all columns prefixed with that lncRNA's identifier, yielding per-subject lncRNA contributions. These were visualised as heatmaps, stacked bar charts, and beeswarm plots to enable subject-level interpretation of which lncRNAs drove each classification decision. 

\begin{figure}[tb]
	\centering
		\includegraphics[width=3.5in]{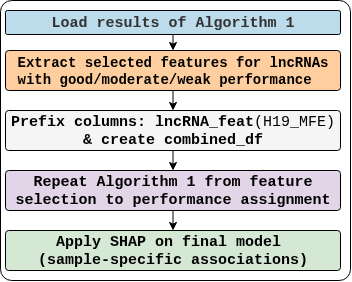}
	\caption{Deatiled flow of the proposed combined model ML framework for finding subject-specific association.}
	\label{fig:ml_pipeline_obj2}
\end{figure}

A detailed pseudo-code representation of the proposed methodology is provided in the Algorithm~\ref{alg:algorithm2}. And detailed flowchart is provided in the Figure \ref{fig:ml_pipeline_obj2}.

Final evaluation was performed using repeated stratified hold-out validation with 20 repeats on the available datasets. The classifier and selected feature set were derived from the prior final model selection stage. 
In each of the 20 repetitions, a stratified test set containing 3 NDC and 3 T2D subjects were randomly selected, 
and the remaining subjects used for training. This produced multiple independent train–test splits under different random seeds.

Predictions from all repetitions were pooled, resulting in a single test-set of 120 (i.e, from 20*(3+3)) predictions used for performance evaluation.
ROC curves were generated from these pooled predictions. For comparison, results from LOOCV, stratified 3-fold CV, and the original hold-out validation setup were also included.

\section{Results and Discussion}
\label{sec:Results}

We evaluated the predictive performance of ten long non-coding RNAs (lncRNAs): MALAT1, MEG3, MIAT, ANRIL, GAS5, KCNQ1OT1, H19, BCYRN1, XIST and HOTAIR, across two independent datasets GSE159984 and GSE164416, in order to identify the population-specific lncNRAs associated with T2D and also to find the subject-specific gene-T2D association levels
by integrating four complementary feature sets: expression, sequence, secondary structure, and All (their combination), alongside their validation via three complementary metrics: Leave-One-Out Cross-Validation (LOOCV) , K-Fold, and Repeated Hold-Out evaluation, as illustrated in Table~\ref{tab:detailed_combined}. Although, LOOCV is a suitable evaluation technique for dataset with small subject size, it is preferred to consider the results from Hold-out and k-fold metrics also, to avoid any misleadings. This framework goes far beyond traditional differential expression analysis, uncovering clear and interpretable associations at both the population and individual subject levels, and ultimately providing actionable biomarker insights.

To contextualize the predictive performance of our multi-modal framework, we computed $Log_2$ Fold-Change ($\log_2$FC) and Cliff’s d on TPM-normalized expression values, along with $N_{\mathrm{mod}}$ and Cliff’s d on sequence-derived features, for all ten lncRNAs as mentioned in Table \ref{tab:discriminability_both}. The ranking of lncRNA models based on the expression-only feature set showed strong concordance with $\log_2$FC and expression based Cliff’s d, and the sequence-based rankings similarly aligned with sequence based Cliff’s d and $N_{\mathrm{mod}}$, indicating that the observed performance was driven by genuine differential signal rather than modeling artifacts. Furthermore, the inclusion of $N_{\mathrm{mod}}$ provided insight into whether discriminative information was concentrated in a few features or distributed across the sequence representation. Overall, the proposed framework demonstrates strong potential for identifying gene-disease associations and offers improved biological interpretability.

\begin{table*}[ht]
	\centering
	\caption{Discriminability of lncRNAs across both cohorts. Cliff's $d$ effect sizes: Neg.\,($0 \leq |d| < 0.147$), Sm.\,($0.147 \leq |d| < 0.330$), Med.\,($0.330 \leq |d| < 0.474$, \colorbox{cliffMedium}{yellow}), Lg.\,($|d| \geq 0.474$, \colorbox{cliffLarge}{green})~\cite{romano_appropriate_2006}. $N_{\mathrm{mod}}$ = number of sequence features with $|d| > 0.3$; shading reflects breadth: \colorbox{cliffMedium}{1-9} and \colorbox{cliffLarge}{$\geq$10}. Bold Cliff's $d$ values indicate $|d| \geq 0.3$. Abbreviations: Neg. (Negative), Sm. (Small), Med. (Medium), Lg. (Large), Interp. (Interpretation).}
	\label{tab:discriminability_both}
	\resizebox{\textwidth}{!}{%
		\begin{tabular}{l rrr rrr c rrr rrr}
			\toprule
			\textbf{lncRNA} & \multicolumn{3}{c}{\textbf{GSE159984 Expression}} & \multicolumn{3}{c}{\textbf{GSE159984 Sequence}} & & \multicolumn{3}{c}{\textbf{GSE164416 Expression}} & \multicolumn{3}{c}{\textbf{GSE164416 Sequence}} \\

			\cmidrule(lr){2-4} \cmidrule(lr){5-7} \cmidrule(lr){9-11} \cmidrule(lr){12-14}
			& $\log_2$FC & Cliff's $d$ & Interp. & Cliff's $d$ & $N_{\mathrm{mod}}$  & Interp. & & $\log_2$FC & Cliff's $d$ & Interp. & Cliff's $d$ & $N_{\mathrm{mod}}$  & Interp. \\
			\midrule
			GAS5 & 0.339 & \textbf{0.433} & \cellcolor{cliffMedium}Med. & \textbf{0.500} & \cellcolor{cliffLarge}28 & \cellcolor{cliffLarge}Lg.  & & -0.010 & -0.094 & Neg. & 0.278 & 0 & Sm.  \\
			MEG3 & -0.334 & -0.148 & Sm. & \textbf{0.447} & \cellcolor{cliffLarge}221 & \cellcolor{cliffMedium}Med.  & & -0.259 & -0.148 & Sm. & \textbf{0.568} & \cellcolor{cliffLarge}78 & \cellcolor{cliffLarge}Lg.  \\
			XIST & -1.307 & \textbf{-0.322} & Sm. & \textbf{0.354} & \cellcolor{cliffLarge}11 & \cellcolor{cliffMedium}Med.  & & 0.002 & 0.054 & Neg. & 0.286 & 0 & Sm.  \\
			MALAT1 & 0.263 & 0.154 & Sm. & \textbf{0.389} & \cellcolor{cliffLarge}14 & \cellcolor{cliffMedium}Med.  & & 0.624 & \textbf{0.544} & \cellcolor{cliffLarge}Lg. & \textbf{0.385} & \cellcolor{cliffLarge}24 & \cellcolor{cliffMedium}Med.  \\
			KCNQ1OT1 & 0.182 & 0.041 & Neg. & \textbf{0.387} & \cellcolor{cliffMedium}6 & \cellcolor{cliffMedium}Med.  & & 0.492 & 0.259 & Sm. & \textbf{0.490} & \cellcolor{cliffLarge}46 & \cellcolor{cliffLarge}Lg.  \\
			ANRIL & 0.240 & 0.109 & Neg. & \textbf{0.318} & \cellcolor{cliffMedium}3 & Sm.  & & 0.488 & 0.168 & Sm. & \textbf{0.410} & \cellcolor{cliffLarge}14 & \cellcolor{cliffMedium}Med.  \\
			BCYRN1 & 0.047 & 0.080 & Neg. & 0.272 & 0 & Sm.  & & 0.422 & 0.202 & Sm. & \textbf{0.379} & \cellcolor{cliffLarge}13 & \cellcolor{cliffMedium}Med.  \\
			MIAT & 0.258 & 0.185 & Sm. & 0.120 & 0 & Neg.  & & 0.476 & 0.125 & Neg. & 0.103 & 0 & Neg.  \\
			H19 & -0.878 & -0.268 & Sm. & 0.015 & 0 & Neg.  & & 1.539 & \textbf{0.308} & Sm. & 0.000 & 0 & Neg.  \\
			HOTAIR & -1.256 & -0.058 & Neg. & 0.000 & 0 & Neg.  & & -1.351 & -0.113 & Neg. & 0.000 & 0 & Neg.  \\
			\bottomrule
			
	\end{tabular}}
\end{table*}

\subsection{Population-Specific T2D-Associated lncRNAs}

\begin{table*}[t]
	\centering
	\small
	\renewcommand{\arraystretch}{1.3}
	\setlength{\tabcolsep}{2pt}
	\caption{Population-specific classification results for associated lncRNAs across GSE159984 and GSE164416 datasets}
	\label{tab:detailed_combined}
	\begin{tabular}{l l l l | c c | c c c c c | c c | c}
		\hline
		\multicolumn{4}{c}{} & \multicolumn{2}{c}{K-Fold} & \multicolumn{5}{c}{LOOCV} & \multicolumn{2}{c}{Hold-out} &  \\
		GSE & lncRNA & Feature & Model & AUC & MCC & AUC & MCC & F1 & Rec & Spec & AUC & MCC & Perf \\
		\hline
		GSE159984 & \textbf{ANRIL} & \textbf{All} & LDA & 0.820000 & 0.406000 & 0.778000 & 0.485000 & 0.632000 & 0.600000 & 0.870000 & 0.828000 & 0.449000 & \tierW \\
		GSE159984 & \textbf{ANRIL} & \textbf{Sequence} & LDA & 0.820000 & 0.406000 & 0.778000 & 0.485000 & 0.632000 & 0.600000 & 0.870000 & 0.828000 & 0.449000 & \tierW \\
		GSE159984 & \textbf{GAS5} & \textbf{All} & LR & 0.818000 & 0.499000 & 0.791000 & 0.441000 & 0.595000 & 0.550000 & 0.870000 & 0.816000 & 0.469000 & \tierW \\
		GSE159984 & \textbf{GAS5} & \textbf{Expression} & ET & 0.689000 & 0.436000 & 0.647000 & 0.290000 & 0.545000 & 0.750000 & 0.565000 & 0.688000 & 0.338000 & \tierW \\
		GSE159984 & \textbf{GAS5} & \textbf{Sequence} & LDA & 0.776000 & 0.365000 & 0.683000 & 0.417000 & 0.605000 & 0.650000 & 0.783000 & 0.758000 & 0.385000 & \tierW \\
		GSE159984 & \textbf{MEG3} & \textbf{All} & ET & 0.775000 & 0.423000 & 0.698000 & 0.391000 & 0.600000 & 0.750000 & 0.674000 & 0.728000 & 0.357000 & \tierW \\
		GSE159984 & \textbf{MEG3} & \textbf{Sequence} & ET & 0.775000 & 0.423000 & 0.698000 & 0.391000 & 0.600000 & 0.750000 & 0.674000 & 0.728000 & 0.357000 & \tierW \\
		GSE159984 & \textbf{XIST} & \textbf{All} & RF & 0.655000 & 0.319000 & 0.627000 & 0.417000 & 0.605000 & 0.650000 & 0.783000 & 0.668000 & 0.311000 & \tierW \\
		GSE159984 & \textbf{XIST} & \textbf{Expression} & ET & 0.692000 & 0.316000 & 0.650000 & 0.318000 & 0.561000 & 0.800000 & 0.543000 & 0.717000 & 0.327000 & \tierW \\
	\hline
		GSE164416 & \textbf{ANRIL} & \textbf{All} & ET & 0.741000 & 0.397000 & 0.741000 & 0.395000 & 0.708000 & 0.590000 & 0.833000 & 0.795000 & 0.386000 & \tierW \\
		GSE164416 & \textbf{ANRIL} & \textbf{Sequence} & ET & 0.741000 & 0.397000 & 0.741000 & 0.395000 & 0.708000 & 0.590000 & 0.833000 & 0.795000 & 0.386000 & \tierW \\
		GSE164416 & \textbf{KCNQ1OT1} & \textbf{All} & KNN & 0.799000 & 0.497000 & 0.804000 & 0.540000 & 0.794000 & 0.692000 & 0.889000 & 0.834000 & 0.561000 & \tierM \\
		GSE164416 & \textbf{KCNQ1OT1} & \textbf{Sequence} & KNN & 0.799000 & 0.497000 & 0.804000 & 0.540000 & 0.794000 & 0.692000 & 0.889000 & 0.834000 & 0.561000 & \tierM \\
		GSE164416 & \textbf{MALAT1} & \textbf{All} & DT & 0.729000 & 0.338000 & 0.685000 & 0.287000 & 0.714000 & 0.641000 & 0.667000 & 0.675000 & 0.319000 & \tierW \\
		GSE164416 & \textbf{MALAT1} & \textbf{Expression} & ET & 0.774000 & 0.466000 & 0.739000 & 0.464000 & 0.789000 & 0.718000 & 0.778000 & 0.767000 & 0.464000 & \tierM \\
		GSE164416 & \textbf{MEG3} & \textbf{All} & LDA & 0.776000 & 0.458000 & 0.682000 & 0.332000 & 0.800000 & 0.821000 & 0.500000 & 0.737000 & 0.387000 & \tierW \\
		GSE164416 & \textbf{MEG3} & \textbf{Sequence} & LDA & 0.776000 & 0.458000 & 0.682000 & 0.332000 & 0.800000 & 0.821000 & 0.500000 & 0.737000 & 0.387000 & \tierW \\
		\hline
	\end{tabular}
	
\end{table*}

Across both datasets, performance varied substantially depending on the lncRNA and feature set employed. Prediction results are displayed in the Table ~\ref{tab:detailed_combined}. (Note: Only the associated per lncRNA feature-set combination results for their best model are displayed in these tables. Selection for the best model was done using the equation \ref{eq:model_score}.)

In \textbf{GSE159984}, lncRNAs \textbf{GAS5, ANRIL, MEG3, and XIST} demonstrated weak association with T2D, rest of the lncRNAs failed to show any associations. 

GAS5 demonstrated weak performance in expression, sequence and all feature-sets configurations based ML models with all feature-set having the best performance among them. Its all feature-set based model displayed k-fold AUC value of 0.81, LOOCV AUC of 0.79 and Hold-out of 0.816, which being highest amongst its feature-sets. Selected features for all feature-set configuration were found to be the combination of selected sequence and expression feature-sets, which might indicate the reason for improvement in overall performance in all feature-set compared to individual expression and sequence feature-set based models. These results for GAS5 expression feature based models, also demonstrated concordance with its expression based Cliff's d and Log2FC values, similarly strong concordance with its sequence features was seen.

MEG3 also demonstrated weak association for its sequence and all feature-set configurations, although all feature-set results were same as sequence feature-set as the selected features were same in both configurations. MEG3 demonstrated AUC values of 0.77, 0.69 and 0.72 under k-fold, LOOCV, hold-out evaluations respectively. Its sequence features based Cliff's d and $N_{mod}$ values of 0.477 and 221 (highest among all lncRNAs) also strengthen the results of its ML model.

Similar to MEG3, ANRIL also demonstrated weak association for sequence and all feature-sets, where all feature-set model again contains same features as in sequence based models. It had a small Cliff’s d effect size and a low $N_{mod}$, indicating that these statistical measures did not capture the association as strongly as captured by the ML model.

XIST demonstrated weak association in both expression and all feature-sets, overall performance of expression feature-set is higher compared to the all feature-set, suggesting the reason for performance drop might be usage of additional features from sequence or structure feature-sets which were not associated. Its Cliff's d value of $-0.322$ was second highest in this population and its $Log_2FC$ value of $-1.307$ suggesting downregulation, indicating chances of some association.

In \textbf{GSE164416}, lncRNAs \textbf{ANRIL, KCNQ1OT1, MALAT1, and MEG3} displayed association ranging from weak to moderate, while the rest of lncRNAs failed to show any association.
KCNQ1OT1 demonstrated moderate associtation for its sequence and all feature-set configurations with AUC values of 0.79, 0.8, and 0.83 for k-fold, LOOCV, and Hold-out respectively. Its performance was best among all the lncRNAs of both datasets. Sequence features were dominant for this lncRNA as both these configurations have same selected sequence features. This important association would have been missed if only expression features were used. These results align with its large sequence's Cliff's d effect size and $N_{mod}$ values of 0.49 and 46 respectively, which further strengthens the ML model results.

MALAT1 displayed moderate association for expression and weak for its all feature-sets based models. The performance in all feature-set dropped slightly compared to that of expression feature based model alone, suggesting other selected features might not be that discriminative. The AUC values of the expression-based model were 0.77, 0.73, and 0.76 across the three validation schemes indicate moderate performance, which was consistent with the expression-based Cliff’s $d$ effect size and $\log_2\mathrm{FC}$ values of 0.544 and 0.624, respectively.

MEG3 also demonstrated weak association for its sequence and all feature-sets configurations. Equal performance for both as the selected features were same. Its sequence features based cliff's d value was highest among all lncRNAs of both datasets, and it had a large value of $N_{mod}$ also, which displayed the concordance with its sequence based ML model performance.

ANRIL demonstrated weak performance for its all and sequence feature-sets, both having same selected features. It had a decent sequence based Cliff's d value of 0.41 and showed moderate interpretability, aligned with the weak associations observed in the ML models.

In both the datasets, MEG3 and ANRIL were common lncRNAs showing weak association for their sequence features. Based upon the results, total 6 lncRNAs i.e., GAS5, MALAT1, MEG3, KCNQ1OT1, XIST and ANRIL out of 10 were found to have some association in atleast one of these datasets, emerging as the most promising candidates for further biomarker investigation. The lncRNAs associated along with their feature modalities in specific populations can be used to further track the root cause of the disease and can help to come up with suitable treatment protocols.

\subsection{Subject-specific associations}

\begin{figure*}[!t]
	\centering
	
	\subfloat[Relative SHAP contribution for GSE159984 T2D cases]{%
   	 \includegraphics{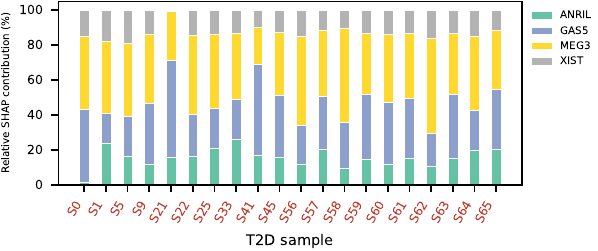}
		\label{fig:shap_gse159984}
	}
	
	\vspace{0cm}
	
	\subfloat[Relative SHAP contribution for GSE164416 T2D cases]{
		\includegraphics[width=\linewidth]{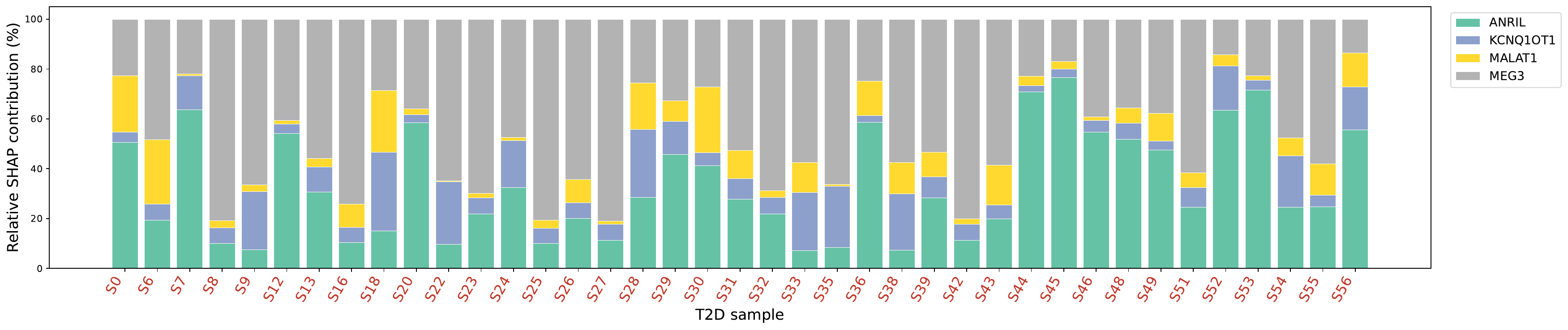}
		\label{fig:shap_gse164416}
	}
	
	\caption{SHAP summary plots for both cohorts showing cohort-specific lncRNA contributions towards T2D association.}
	\label{fig:shap_combined}
\end{figure*}

\begin{table}[htbp]
	\centering
	\caption{Performance of final model. KF\,=\,K-fold, LOO\,=\,Leave-One-Out, HO\,=\,Hold-out, FT\,=\,Final-test.}
	\label{tab:final_model_results}
	\footnotesize
	\setlength{\tabcolsep}{3pt}
	\setlength{\abovecaptionskip}{0pt}
	\setlength{\belowcaptionskip}{0pt}
	\renewcommand{\arraystretch}{1}
	\begin{tabular}{llccccc}
		\toprule
		\textbf{Dataset} & \textbf{Model} & \textbf{Metric} & \textbf{KF} & \textbf{LOO} & \textbf{HO} & \textbf{FT} \\
		\midrule
		GSE159984 & RF & AUC & 0.788 & 0.749 & 0.786 & 0.69 \\
		& & MCC & 0.509 & 0.516 & 0.520 & 0.49 \\
		& & Acc & 0.788 & 0.803 & 0.801 & 0.73 \\
		& & Rec & 0.651 & 0.600 & 0.640 & 0.58 \\
		& & Spec & 0.850 & 0.891 & 0.866 & 0.88 \\
		& & F1 & 0.651 & 0.649 & 0.640 & 0.69 \\
		& & \textbf{Perf} & \multicolumn{3}{c}{\cellcolor{yellow!30}Moderate} & \tierW \\
		\addlinespace
		GSE164416 & ET & AUC & 0.857 & 0.823 & 0.862 & 0.77 \\
		& & MCC & 0.626 & 0.624 & 0.638 & 0.50 \\
		& & Acc & 0.825 & 0.825 & 0.815 & 0.75 \\
		& & Rec & 0.821 & 0.821 & 0.794 & 0.72 \\
		& & Spec & 0.833 & 0.833 & 0.858 & 0.78 \\
		& & F1 & 0.864 & 0.865 & 0.845 & 0.74 \\
		& & \textbf{Perf} & \multicolumn{3}{c}{\cellcolor{green!20}\textbf{Good}} & \tierM \\
		\addlinespace
		\bottomrule
	\end{tabular}
\end{table}

\begin{figure}[tb]
	\centering
	
	\begin{minipage}{0.5\textwidth}
		\centering
		\includegraphics[width=3.5in]{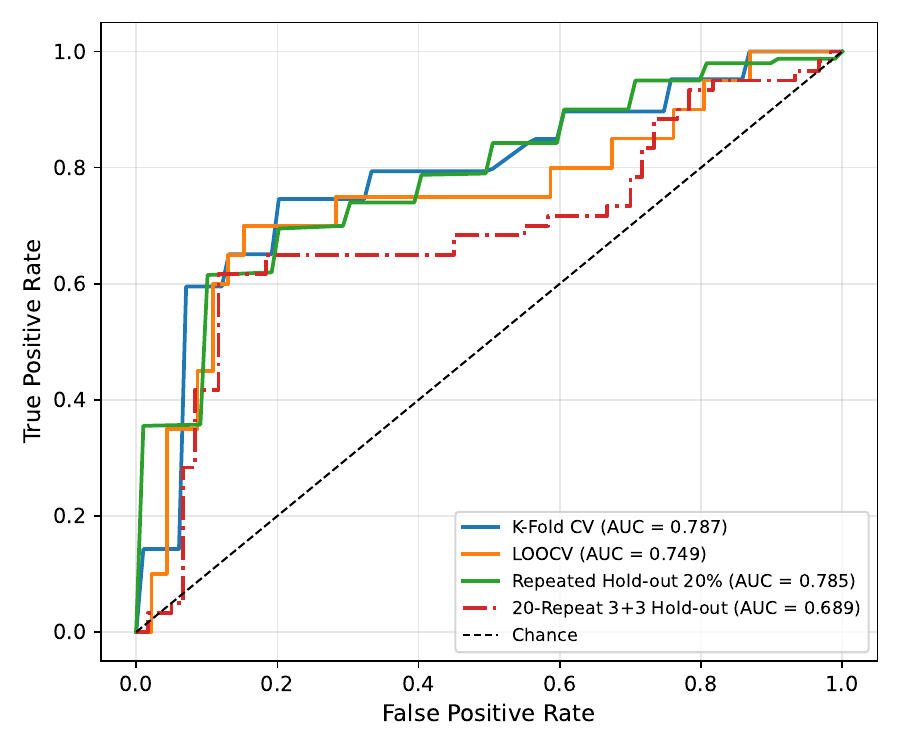}
		\text{(A)} ROC curves for the final model in the GSE159984 
	\end{minipage}
	\hfill
	\begin{minipage}{0.5\textwidth}
		\centering
		\includegraphics[width=3.5in]{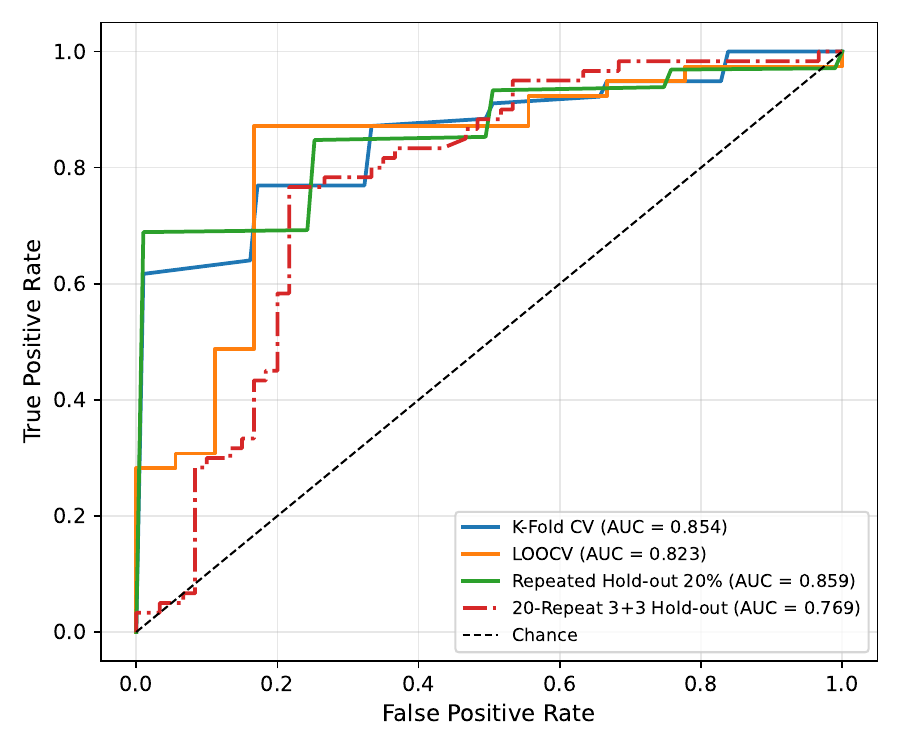}
		\text{(B)} ROC curves for the final model in the GSE164416
	\end{minipage}
	
	\caption{%
		ROC curves for the final lncRNA models across both cohorts. 
	}
	\label{fig:roc_combined}
\end{figure}

(a) Integrated multi-lncRNA model for GSE159984: \\

The final integrated model was selected from the eight classifiers trained on the selected features of associated lncRNAs. We found that on this combined dataset RF classifier was the dominant base learner, demonstrating moderate tier performance which was highest compared to the individual lncRNAs models. Final model was further used for detailed evaluation and SHAP values were computed.
Performance demonstrated by this final model was as follows (Table \ref{tab:final_model_results}):
Under the k-fold scheme, the integrated model achieved moderate AUC of 0.78, MCC of 0.5, and moderate performance for other metrics too. In $LOOCV$ scheme also it achieved moderate AUC of 0.74 and 0.78 in Hold-out scheme, with decent performance in all other metrics too. Overall, this integrated model's performance improved a lot in comparison to the individual lncRNA based models, suggesting the combined association roles for GAS5, MEG3, ANRIL, and XIST features in this population.

SHAP-based interpretability analysis of this integrated model
revealed that MEG3 was the dominant contributor to individual subject
predictions, exhibiting the largest SHAP values across majority of subjects, with GAS5 also having very close values to MEG3, as depicted in the Figure \ref{fig:shap_gse159984}. \\

(b) Integrated multi-lncRNA model for GSE164416: \\
For the final model, ET classifier was the dominant base learner in this population, demonstrating good tier performance, which was again highest compared to the individual lncRNAs models. 
Performance demonstrated by this final model was as follows (Table \ref{tab:final_model_results}):
Under the k-fold scheme, the integrated model achieved a good AUC of 0.85, MCC of 0.62, and good performance for other metrics too. In $LOOCV$ scheme also it achieved good AUC of 0.82 and 0.86 in Hold-out scheme, with good performance in all other metrics too. Overall, this integrated model's performance improved by significant amount with respect to the individual lncRNAs features based models alone in this population also.

SHAP-based interpretability analysis of GSE164416 revealed MEG3 was the dominant contributor to individual subject predictions, exhibiting the largest SHAP values across the majority of subjects as depicted in the Figure \ref{fig:shap_gse164416}.

In both datasets, combining associated lncRNAs features led to a clear improvement in performance metrics, suggesting that complementary disease signals across lncRNAs can enhance generalization.

Performance metrics were evaluated for final models of both datasets (see Table \ref{tab:final_model_results}) by performing final-test evaluation where 3 samples from each class were held out completely from the training dataset of final model and only used for evaluation, this process was repeated 20 times, and finally metrics were evaluated using their pooled results. In GSE159984, performance observed on final test was decent with AUC value of 0.69, MCC of 0.49, aligning with moderate tier performance of its final model but the effect of class imbalance can be seen from its weak recall value of 0.58 suggesting that model predicts some T2D cases as NDC too which was due to NDC being majority class, whereas its specificity value of 0.88 was good. In GSE164416, performance on final test also seems good enough for such imbalanced dataset, with AUC value of 0.77, MCC of 0.5 and good value for both recall (0.72) and specificity (0.78). This further substantiate the robustness of this model prediction even on such a small dataset.

The ROC analysis (Figure \ref{fig:roc_combined}) demonstrates consistent model behaviour across both datasets. For GSE159984, the LOOCV-based evaluation yielded an AUC of 0.749, closely matching the repeated hold-out AUC of 0.785 and 3-fold CV AUC of 0.787, indicating moderate and stable discriminative performance across validation schemes. The final test AUC of 0.689 displayed slight drop compared to the validation AUC values of full dataset but still demonstrates a decent discriminative power. GSE164416 model showed good performance, with a LOOCV AUC of 0.823, repeated hold-out AUC of 0.859, and 3-fold CV AUC of 0.854, suggesting improved classification and robustness. Its final-test AUC of 0.769 is also close and demonstrates moderate predictive performance.

Despite methodological differences among validation schemes, the results remain closely aligned, indicating stable generalisation behaviour without significant optimism bias. The final-test ROC further reflects performance on independent held-out samples, complementing the validation results. 

Overall, the consistency across all ROC curves reinforces the robustness of the final model, with GSE164416 consistently outperforming GSE159984 across all evaluation protocols. It could be due to the T2D being majority class in GSE164416 whereas T2D was minority in the GSE159984. 
These results suggests that the final model trained on dataset with improved class balance, can potentially be extended to predict disease associations with more accuracy.

\subsection{Comparison with previous studies}
Previous studies are not that closely related to the tasks accomplished in this study, as 
most of them do the classification between T2D and control only, by utilising the clinical parameters mainly ~\cite{nuthakki_machine_2024} or a mixture of clinical and a few common/preset of molecular parameters ~\cite{matboli_machine_2025}. But they don't have capability to answer the predictions made by our framework such as population and subject specific gene-T2D association levels, and their feature set responsible for association.
Compared to recent proteomic approaches employing Olink proximity extension 
assays combined with clinical metadata~\cite{villikudathil_exploring_2024}, our RNA-seq-only 
framework offers complementary advantages. RNA-seq is more accessible and cost-effective than targeted protein quantification platforms, broadening applicability to large existing biobanks. lncRNAs act upstream of protein translation, placing our identified associations closer to the regulatory origin of T2D-associated transcriptional 
programmes. Protein-based approaches cannot derive sequence variant information as done here. 

Advantages of proposed framework over existing methods are as follows.
From population-level associations to subject-specific risk prediction; traditional differential expression methods, such as log$_2$FC and Cliff’s $d$, are widely used to identify genes associated with a disease at the population level. However, these approaches do not provide insights at the level of individual subjects. In contrast, machine learning models are designed for prediction but often lack clear biological interpretability. The proposed framework brings these two perspectives together. It not only enables prediction of T2D risk for individual subjects based on their feature profiles, but also helps uncover population-specific lncRNA-disease associations. This combination offers both practical predictive capability and meaningful biological insights, making it more suitable for precision medicine applications.
 Multi-modal integration and interpretability by jointly using expression, secondary-structure, and sequence-derived features; the framework captures complementary biological signals, with sequence features demonstrating independent predictive value. These representations support ML-based lncRNA-disease association prediction, biomarker identification, and can be further enriched with additional biological context (e.g., structural or regulatory information) to improve mechanistic understanding.
 Subject-specific lncRNA dominance analysis provides sample-level interpretability by quantifying the relative contribution of each lncRNA to individual predictions. This enables identification of subject-specific dominant lncRNAs, offering insights into heterogeneous disease mechanisms and supporting more personalised biological interpretation.
 Variant-level resolution from a single assay using RNA-seq–derived variants, enabling an eQTL-like genotype–phenotype link without Whole-Genome Sequencing (WGS) or Whole-Exome Sequencing (WES).

\section{Conclusion}\label{sec:conclusion}
We investigated ten lncRNAs known to be associated with T2D in two datasets, and identified four associated lncRNAs in each dataset: GAS5, MEG3, XIST, and ANRIL in one dataset, and MALAT1, KCNQ1OT1, ANRIL, and MEG3 in the other. GAS5, MALAT1 and XIST association are primarily expression driven, and as for KCNQ1OT1, MEG3, and ANRIL, their associations are driven primarily by sequnce-level variations only.MALAT1 and KCNQ1OT1 individually showed moderate performance, ranking highest among the identified lncRNAs. Apart from these 6 lncRNAs, remaining 4 failed to show any association in these cohorts. The ML models validation results aligned with statistical analyses (Cliff’s d), supporting biological plausibility, while highlighting the ML framework’s ability to integrate and prioritize multi-feature signals and identify disease associations in unseen subjects.
The results from this proposed framework suggests that it can be used to find the gene-disease association in any particular population cohort and can return interpretable gene-disease association levels unique to subjects of that population. Our framework also distinguishes whether a given lncRNA-T2D association is driven primarily by 
expression-derived features or sequence-structural properties, thereby providing insight into how each candidate association operates and prioritizing genes for targeted functional investigation.

Several limitations should be acknowledged. First, the subject sizes
in both datasets~(GSE159984: n = 66; GSE164416: n = 57) are
modest and class imbalance is also present, which may affect classifier performance and generalisation. Future work should validate the top candidates in larger cohorts with class balance.
Second, sequence features were derived using variants identified directly from RNA-seq data rather than dedicated whole-genome or exome sequencing. As a result, variants in highly expressed lncRNAs are more likely to be captured, while low-abundance variants may be missed, limiting comprehensive patient-specific sequence characterization.

This work can further be extended to find the novel genes association verification and the proposed framework can be positioned as a foundation for precision-medicine approaches in which patient stratification is informed by lncRNAs expression and sequential signatures. Integration of our framework with multi-omics layers, could resolve the upstream regulatory events that initiate lncRNA 
dysregulation in $\beta$-cell failure. Finally, the framework can be extended to other diseases, provided RNA-sequencing data are available, to evaluate its generalizability beyond the T2D context.

\bibliographystyle{IEEEtran}
\bibliography{1st_paper}

\vfill

\end{document}